\documentclass[prb,twocolumn,showpacs,amsmath,amssymb,superscriptaddress]{revtex4-1}
\usepackage{dcolumn}
\usepackage{bm,graphicx}
\usepackage{etoolbox}
\usepackage{url}
\usepackage[colorlinks]{hyperref}
\usepackage{breakurl}
\usepackage{color}

\begin{document}

\title{The energy scale of Dirac electrons in Cd$_3$As$_2$}

\author{M.~Hakl}\email{michaelhakl@email.cz}
\affiliation{Laboratoire National des Champs Magn\'etiques Intenses, CNRS-UGA-UPS-INSA-EMFL, 25, avenue des Martyrs, 38042 Grenoble, France}
\author{S.~Tchoumakov}
\affiliation{LPS, Univ. Paris-Sud, Univ. Paris-Saclay, CNRS UMR 8502, 91405 Orsay, France}
\author{I.~Crassee}
\affiliation{Laboratoire National des Champs Magn\'etiques Intenses, CNRS-UGA-UPS-INSA-EMFL, 25, avenue des Martyrs, 38042 Grenoble, France}
\author{A.~Akrap}
\affiliation{DQMP, University of Geneva, 1211 Geneva 4, Switzerland}
\author{B.~A.~Piot}
\affiliation{Laboratoire National des Champs Magn\'etiques Intenses, CNRS-UGA-UPS-INSA-EMFL, 25, avenue des Martyrs, 38042 Grenoble, France}
\author{C.~Faugeras}
\affiliation{Laboratoire National des Champs Magn\'etiques Intenses, CNRS-UGA-UPS-INSA-EMFL, 25, avenue des Martyrs, 38042 Grenoble, France}
\author{G.~Martinez}
\affiliation{Laboratoire National des Champs Magn\'etiques Intenses, CNRS-UGA-UPS-INSA-EMFL, 25, avenue des Martyrs, 38042 Grenoble, France}
\author{A.~Nateprov}
\affiliation{Institute of Applied Physics, Academy of Sciences of Moldova, 2028 Chisinau, Moldova}
\author{E.~Arushanov}
\affiliation{Institute of Applied Physics, Academy of Sciences of Moldova, 2028 Chisinau, Moldova}
\author{F.~Teppe}
\affiliation{Laboratoire Charles Coulomb, CNRS, Universit\'{e} Montpellier, 34095 Montpellier, France}
\author{R.~Sankar}
\affiliation{Institute of Physics, Academia Sinica, Nankang, 11529 Taipei, Taiwan}
\affiliation{Center for Condensed Matter Sciences, National Taiwan University, Taipei 10617, Taiwan}
\author{Wei-li~Lee}
\affiliation{Institute of Physics, Academia Sinica, Nankang, 11529 Taipei, Taiwan}
\author{J.~Debray}
\affiliation{Universit\'{e} Grenoble Alpes, Institut NEEL, F-38000 Grenoble, France}
\affiliation{CNRS, Institut NEEL, F-38000 Grenoble, France}
\author{O.~Caha}
\affiliation{CEITEC MU and Faculty of Science, Masaryk University, 61137 Brno, Czech Republic}
\author{J.~Nov\'{a}k}
\affiliation{CEITEC MU and Faculty of Science, Masaryk University, 61137 Brno, Czech Republic}
\author{M.~O.~Goerbig}
\affiliation{LPS, Univ. Paris-Sud, Univ. Paris-Saclay, CNRS UMR 8502, 91405 Orsay, France}
\author{M.~Potemski}
\affiliation{Laboratoire National des Champs Magn\'etiques Intenses, CNRS-UGA-UPS-INSA-EMFL, 25, avenue des Martyrs, 38042 Grenoble, France}
\author{M.~Orlita}\email{milan.orlita@lncmi.cnrs.fr}
\affiliation{Laboratoire National des Champs Magn\'etiques Intenses, CNRS-UGA-UPS-INSA-EMFL, 25, avenue des Martyrs, 38042 Grenoble, France}
\affiliation{Institute of Physics, Charles University, Ke Karlovu 5, 12116 Praha 2, Czech Republic}

\date{\today}

\begin{abstract} Cadmium arsenide (Cd$_3$As$_2$) has recently became conspicuous in solid-state physics
due to several reports proposing that it hosts a pair of symmetry-protected 3D Dirac cones.
Despite vast investigations, a solid experimental insight into the band structure of this material is still
missing. Here we fill one of the existing gaps in our understanding of Cd$_3$As$_2$, and based on our Landau level spectroscopy study, we provide an estimate
for the energy scale of 3D Dirac electrons in this system. We find that the appearance of such
charge carriers is limited -- contrary to a widespread belief in the solid-state community -- to a relatively small energy scale (below 40 meV).
\end{abstract}

\pacs{78.20.Ls, 71.28.+d, 71.70.Di}
\maketitle

\section{Introduction}

The presence of three-dimensional (3D) massless Dirac electrons in cadmium arsenide (Cd$_3$As$_2$) is nowadays taken as granted,
from both experimental and theoretical point of views.\cite{WangPRB13,BorisenkoPRL14,NeupaneNatureComm14,LiuScience14} Cd$_3$As$_2$ is thus often cited as
the first identified symmetry-protected 3D Dirac semimetal that is stable under ambient conditions. The cartoon picture of its band structure -- with two Dirac cones located along the tetragonal axis -- is even viewed as a textbook example for this class of materials.

In reality, the band structure of Cd$_3$As$_2$ is far more complex and the presence of Dirac electrons, which are of key interest in this material,\cite{WangPRB13}
is not proven. The broadly extended conical band reported in recent experiments on Cd$_3$As$_2$\cite{BorisenkoPRL14,NeupaneNatureComm14,LiuScience14}
is most likely not related to any Dirac physics.\cite{JeonNatureMater14,AkrapPRL16} Instead, this conical band seems to be a simple consequence of a vanishing band gap in the
material studied.
The low-energy electronic states -- where a pair of truly Dirac cones may indeed be expected\cite{WangPRB13,ConteSR17} -- have been so far barely addressed in experiments.
Hence, at present, the Dirac electrons in Cd$_3$As$_2$ do not represent more than an appealing theoretical construct.

\begin{figure*}
	\centering
	\includegraphics[width=0.6\textwidth,trim={1cm 1cm 3cm 1cm}]{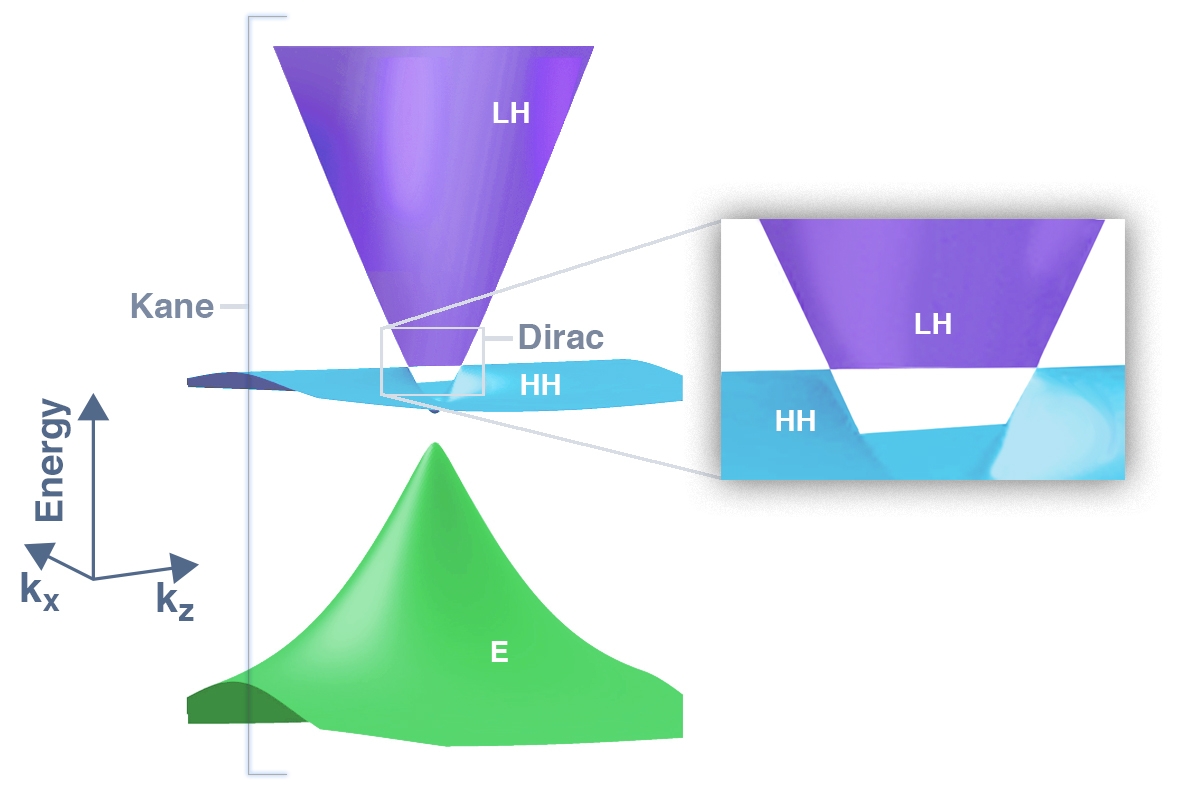}
	\caption{A schematic view of electronic bands in Cd$_3$As$_2$ at the Brillouin zone center. Three electronic
    bands form two types of 3D conical structures: a single cone hosting Kane electrons at the larger energy scale, appearing due to the vanishing
    band gap, and two highly tilted and anisotropic 3D Dirac cones at low energies. The Dirac cones emerge at the crossing points of
    $p$-type bands, which in semiconductors with non-inverted ordering of bands are referred to as heavy (HH) and light (LH)
    hole bands~\cite{CardonaYu}. In materials with an inverted band gap (\emph{e.g.}, in HgTe or Cd$_3$As$_2$), the LH band becomes the conduction band.
        \label{Scheme}}
\end{figure*}

A simplified model of the complex band structure of Cd$_3$As$_2$ has been in the past proposed by Bodnar,\cite{Bodnar77} who treated this material --
in the very first approach -- as a conventional narrow-gap Kane semiconductor/semimetal\cite{KaneJPCS57} with a nearly vanishing band gap ($E_g$).
The band structure of Cd$_3$As$_2$ thus displays, similar to gapless HgCdTe,\cite{OrlitaNaturePhys14,TeppeNatureComm16} a widely extended conical band, which is
centered around the $\Gamma$ point (Fig.~\ref{Scheme}). This cone appears due to the approximate accidental degeneracy of $p$- and $s$-like states of
arsenic and cadmium, respectively, and is not protected by any symmetry.

A closer look at the band structure of Cd$_3$As$_2$ implied by the model of Bodnar also reveals symmetry-protected Dirac cones. These highly-anisotropic and strongly tilted
cones appear -- or at least, they are theoretically expected -- at low energies, around the crossing points of two arsenic-like $p$-type bands (LH and HH bands in Fig.~\ref{Scheme}).
Such crossing points emerge when the cubic symmetry of a Kane semiconductor/semimetal is reduced to a tetragonal one,
which is the case of Cd$_3$As$_2$. The impact of the symmetry lowering on the band structure is quantified by crystal field splitting $\delta$.\cite{KildalPRB74}
This parameter describes the energy separation of two $p$-type bands at $k=0$ (degenerate in cubic semiconductors\cite{CardonaYu})
and directly measures the energy scale of Dirac electrons in Cd$_3$As$_2$.

Reliable experimental estimates of the crystal field splitting and band gap parameters -- those which are directly determining the shape
and scale of Dirac cones in Cd$_3$As$_2$ -- have been missing so far. In this paper, we examine the Cd$_3$As$_2$ bands using Landau level spectroscopy and fill this
gap in our understanding of this material. Our results show that cadmium arsenide is a semimetal with a small inverted gap
of $E_g=-(70\pm20)$~meV, which may host three-dimensional symmetry-protected Dirac electrons, but only at the energy scale not exceeding several dozens of meV.

\section{Experimental results and discussion}

The experiment was performed on a free-standing 80-$\mu$m-thick slab of $\mbox{Cd}_3\mbox{As}_2$ with lateral dimensions of $2\times1$~mm$^2$, which  was grown in a downstream
of the tubular furnace from the polycrystalline phase and later detached from the substrate.\cite{RamboCJP79} The $x$-ray measurements provided us with
a diffraction pattern characteristic of Cd$_3$As$_2$\cite{AliIC14} and also showed the existence of a few monocrystalline grains -- all with the
(112)-orientation, which is the typical growth direction of this material. Similar to other Cd$_3$As$_2$ crystals, the sample shows $n$-type conductivity\cite{RosenmanJPCS69,ZdanowiczTSF79,ArushanovPCGC80,ArushanovPCGC92} due to intrinsic defects, with an electron concentration slightly
below $10^{19}$~cm$^{-3}$.

\begin{figure}
	\centering
	\includegraphics[width=0.47\textwidth]{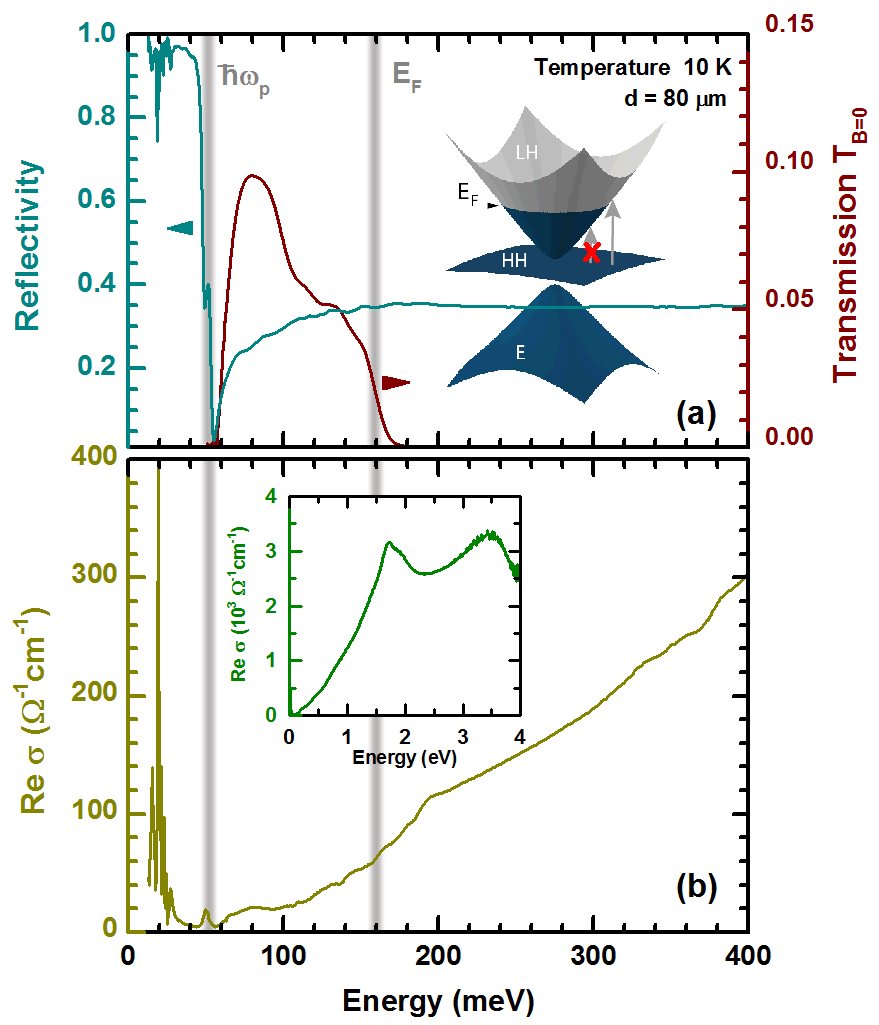}
	\caption{(a) Transmission and reflectivity spectra of the investigated thin layer of Cd$_3$As$_2$ in the far and middle infrared spectral ranges. The upper and lower limits of the transmission window are given by the plasma energy $\hbar\omega_p$ and by the onset of interband absorption, defined by the particular position of the Fermi energy $E_F$ (by Pauli blocking). This is shown in the inset of (a), where the band structure expected within the (gapped) Kane model is schematically plotted. Spectrally sharp features in the reflectivity spectrum visible at low energies are due to infrared active optical phonon modes. The optical conductivity spectrum deduced using Kramers-Kronig relations from the reflectivity curve is plotted in (b). The observed broadening of the onset of interband absorption indicates somewhat inhomogeneous distribution of electrons across the sample, with the mean value of the Fermi energy $E_F\approx 160$~meV. The optical conductivity in a broader spectral range is plotted in the inset of (b).
		\label{Zero-field}}
\end{figure}

The studied sample was characterized by zero-field optical spectroscopy in both transmission and reflectivity configurations (Fig.~\ref{Zero-field}a). The reflectivity spectrum, measured using an infrared microscope combined with a conventional Fourier-transform spectrometer, indicates the presence of free charge carriers by a characteristic plasma edge at $\hbar\omega_p \approx 50$~meV. The fine structure of the plasma edge was also observed in other studies, see, \emph{e.g.}, Refs.~\onlinecite{GeltenSSC80,NeubauerPRB16,AkrapPRL16,CrasseePRB18} and it is likely related to non-homogeneous (lateral as well as perpendicular to the surface) distribution of electrons in the explored sample.

\begin{figure*}
	\centering
	\includegraphics[width=0.69\textwidth]{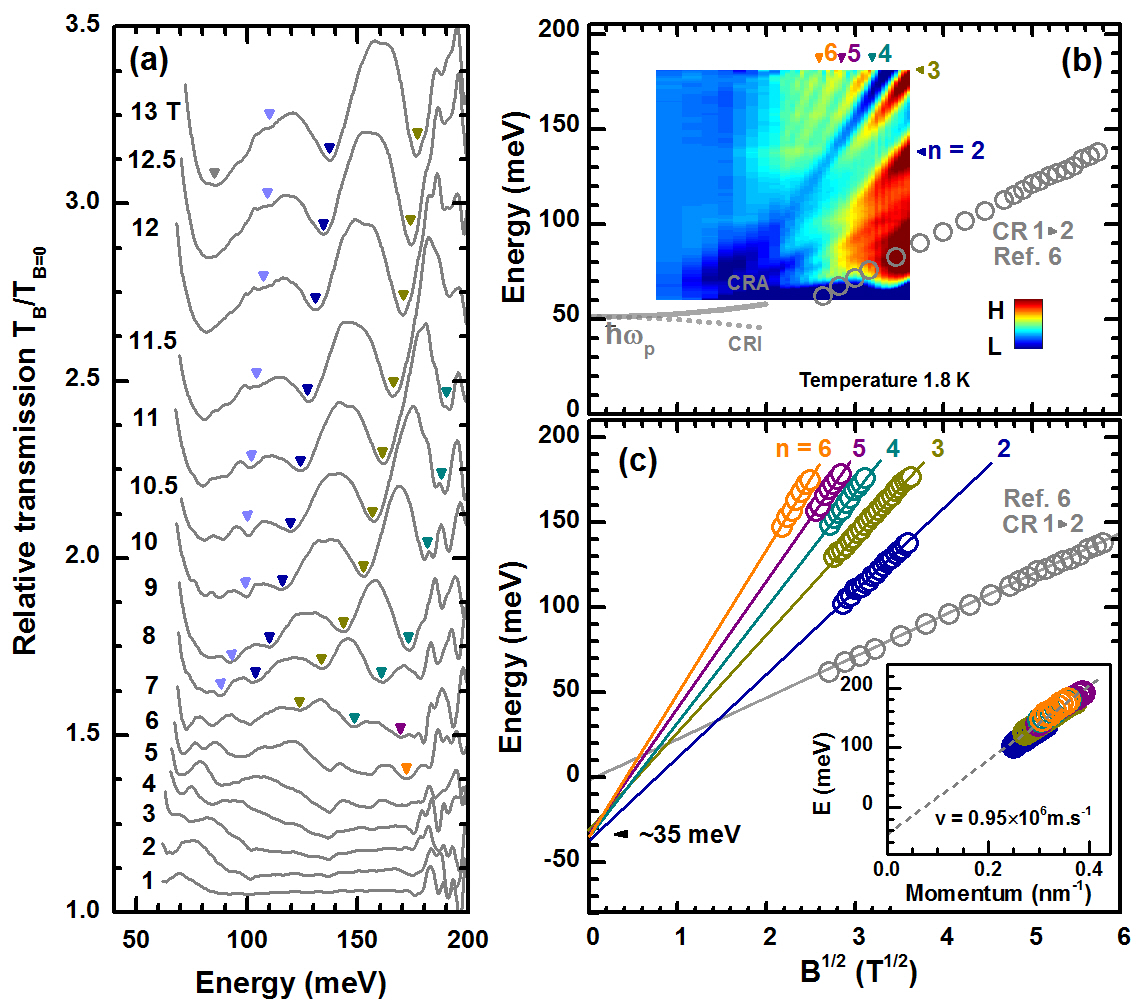}
	\caption{(a) Relative magneto-transmission spectra $T_B/T_{B=0}$ of the explored Cd$_3$As$_2$ sample plotted for selected values of the magnetic field $B$ applied. Individual
                 inter-LL resonances are denoted by triangles.
             (b) The false color-map of magneto-absorbance spectra. The gray circles show the cyclotron energy (fundamental mode)
                  obtained in the previous high-field magneto-reflectivity experiments~\cite{AkrapPRL16}.
             (c) The experimentally extracted energies of inter-LL resonances versus $\sqrt{B}$. To exclude effects related to the coupling of interband (single-particle)
               excitations with plasmon (so-called Bernstein modes), only resonances at photon energies above $\hbar\omega=100$~meV are plotted ($\hbar\omega_p\approx 50$~meV).
               The inset shows the linear energy-momentum dependence (the separation of the conduction and valence band) obtained using a simple semi-classical
               model based on the Bohr-Sommerfeld quantization, see the text.
                \label{Fan chart}}
\end{figure*}

The transmission spectrum of the sample shows a well-defined transparency window, which is marked by vertical gray bars in~Fig.~\ref{Zero-field}a. At low energies,
this transparency window opens just above the plasma edge. At high energies, the transmission window closes due to the onset of interband absorption. The dominant contribution
to this absorption arises from electrons excited from the flat (HH) band to the partially occupied upper conical band,\cite{AkrapPRL16} as schematically shown in the inset of Fig.~\ref{Zero-field}a. The position of the onset of the interband absorption is thus determined by the particular position of the Fermi energy (due to Pauli blocking),
which provides us with an estimate of $E_F\approx160$~meV. In this estimate we have completely neglected the dispersion of the flat band, which is in reality a hole-like, approximately parabolic band. The corresponding effective mass should be comparable with heavy hole masses in conventional semiconductors\cite{CardonaYu} (\emph{i.e.}, approaching the mass of a bare electron in vacuum). The dispersion of this band thus may be neglected in the close vicinity of the  $\Gamma$ point -- where the studied (magneto-)optical excitations originate -- but has to be considered for momenta comparable with the size of the Brillouin zone, which is the typical scale of ARPES experiments.\cite{BorisenkoPRL14,NeupaneNatureComm14,LiuScience14}

The onset of interband excitations may be directly seen in the optical conductivity spectrum (Fig.~\ref{Zero-field}b), which was obtained using the standard
Kramers-Kronig analysis of the reflectivity response, complemented by ellipsometry data at high photon energies (inset of Fig.~\ref{Zero-field}b).
The smooth spectral profile of this onset indicates a certain spread of the Fermi energy across the probed sample, but likely also some defect/impurity-related absorption
within the optical band gap. Above the onset of interband excitations, the optical conductivity follows a linear, or more precisely a weakly superlinear, dependence
on the photon energy, which is in perfect agreement with previous studies.\cite{NeubauerPRB16,AkrapPRL16} The optical conductivity increasing linearly with the photon frequency, $\sigma(\omega)\propto \omega$, is typical of systems with 3D massless electrons.\cite{TimuskPRB13,OrlitaNaturePhys14} The strong increase in the real part of the
optical conductivity at low energies is related to intraband (Drude-type, free-carrier) excitations.

\begin{figure*}
	\centering
	\includegraphics[width=0.7\textwidth,trim={0cm 0cm 0cm 0cm}]{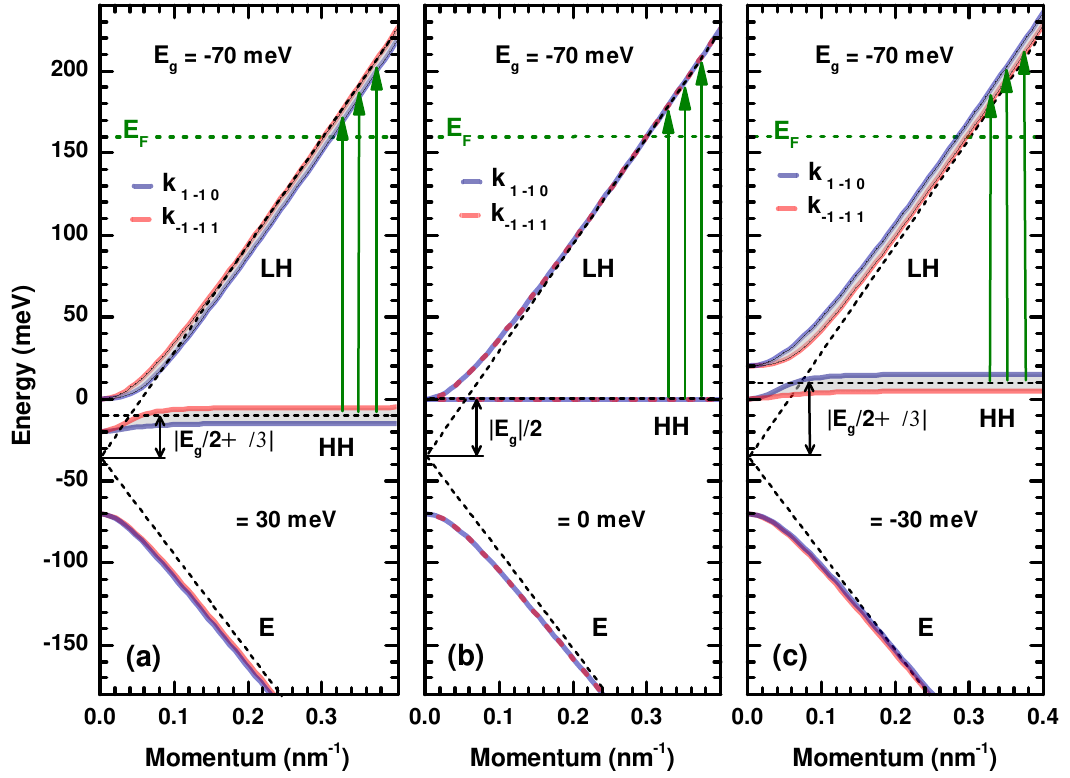}
	\caption{The electronic band structure of Cd$_3$As$_2$ calculated within the Bodnar/Kane model for the fixed values of $E_g=-70$~meV and $v=0.94\times10^6$~m/s, for three different
              strengths of crystal field splitting: $\delta=30, 0$ and -30~meV, in parts (a), (b) and (c) respectively. The Dirac cones are expected to appear for positive $\delta$, while
              for the negative one, the system should enter a topologically insulating phase.
              The electronic bands have been plotted along (1 -1 0) and (-1 -1 1)
              directions, which are perpendicular to the applied magnetic field, $B\|(112)$, as well as to each other. The non-zero $\delta$ parameter mainly impacts the HH band, which becomes modulated, anisotropic and shifted in energy by the average amount of $\delta/3$. As shown by the corresponding asymptotes, the upper cone thus becomes shifted with respect to the flat band, implying thus the offset of $E_g/2+\delta/3$ for interband absorption (and interband inter-LL excitations in Fig.~\ref{Fan chart}c).
              The vertical arrows denote interband excitations allowed by the occupation of the upper conical band, which dominantly contribute to interband absorption.
		\label{Energy-momentum}}
\end{figure*}

The existence of the well-defined transparency window allowed us to probe the response of the sample in magneto-transmission
experiments. To perform such measurements, the radiation of a globar was analysed by a Fourier transform spectrometer and using light-pipe optics, delivered to the sample placed in a superconducting coil. The transmitted light was detected by a composite bolometer, placed directly below the sample and kept at the same temperature as the sample (1.8~K).
The measurement was done in the Faraday geometry with light propagating perpendicular to the (112) crystal face.

A series of excitations, at energies monotonically increasing with $B$, is clearly seen in the relative magneto-transmission
spectra $T_B/T_{B=0}$ (Fig.~\ref{Fan chart}a) as well as in the false color plot of relative magneto-absorbance $-\ln[T_B/T_{B=0}]$ (Fig.~\ref{Fan chart}a).
These excitations appear in the spectra at photon energies significantly higher than the cyclotron energy in the quantum limit, deduced
from our preceding high-field magneto-reflectivity experiment.\cite{AkrapPRL16} This points towards the interband origin of these inter-Landau level (inter-LL) excitations.

In the limit of low magnetic fields, these interband resonances seem to asymptotically approach the plasma energy (see color plot in Fig.~\ref{Fan chart}b). This behavior
may be viewed as an avoided crossing of interband inter-LL excitations with a plasma resonance, known as the Bernstein modes \cite{bernstein_original}.
Since the sample is fully opaque at photon frequencies below $\omega_p$, we only observe the upper branch of these coupled modes. Such coupling
can be clearly visualized in 2D structures with an additional lateral periodic potential. For instance, such modes have been observed in magneto-optics of quantum wells\cite{bernstein_2d_exp1,bernstein_2d_exp2} with the use
of surface gratings in order to match inter-LL and plasma frequencies. In 3D samples, such as here, this coupling can be observed more easily since the
plasma frequency does not vanish at ${\bf q} = 0$ and it was already demonstrated in Raman spectroscopy of $n$-type GaAs.\cite{bernstein_3d_exp}
The magneto-optical effects due to this coupling may be richer in systems with a linear dispersion, the gapless character of which implies a series
of interband inter-LL excitations at relatively low energies, as discussed in the context of graphene.\cite{bernstein_graphene} To compare the energy
scales of the avoided crossing related to Bernstein modes with the field-induced splitting of the plasma edge expected within the classical theory of magneto-plasma,\cite{PalikRPP70}
we plotted in Fig.~\ref{Fan chart}b the theoretical positions of the cyclotron resonance active and inactive (CRA and CRI) modes for the corresponding
cyclotron mass $m_c=E_F/v^2$.

At high photon energies, well above the plasma energy ($\omega\gg\omega_p$), the observed transitions correspond to single-particle interband
excitations between Landau levels. Tracing their field-dependence may provide us with a useful insight into the
electronic band structure of Cd$_3$As$_2$. All excitations show a well-defined $\sqrt{B}$ dependence, which is in general characteristic of massless charge carriers,
nevertheless, with a well defined negative offset of about $-(35\pm10)$~meV (Fig.~\ref{Fan chart}c). This is in contrast to the cyclotron energy in the quantum limit,
which is also linear in $\sqrt{B}$, but without any offset.\cite{AkrapPRL16}

\begin{figure*}
	\centering
	\includegraphics[width=0.73\textwidth]{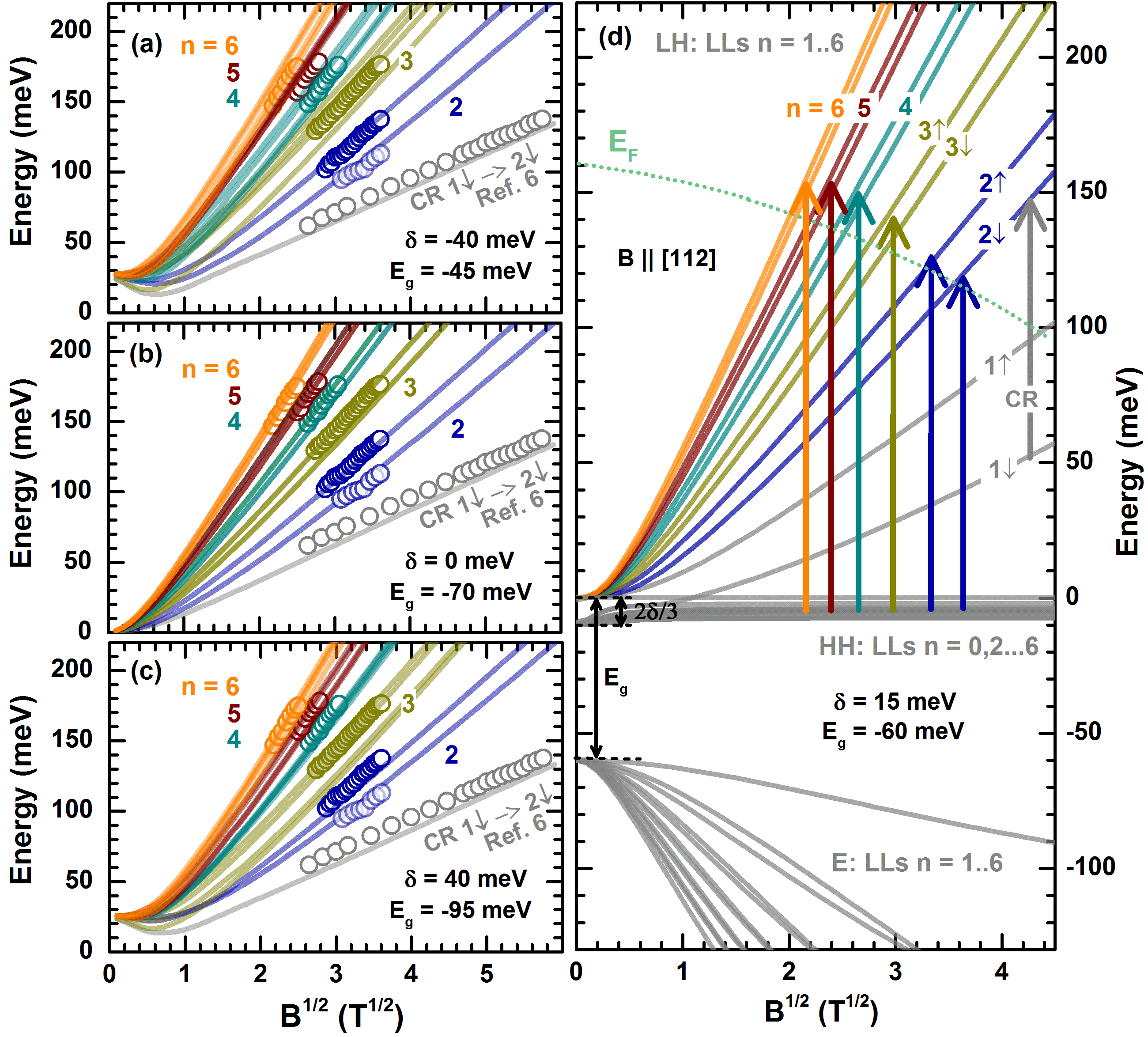}
	\caption{Parts (a-c): The experimentally determined fan chart of observed inter-LL excitations compared to theoretically calculated excitation energies
     within the Bodnar (Kane) model for three different combinations of $E_g$ and $\delta$, which fulfill the condition $E_g/2+\delta/3\approx-35$~meV implied
     by the offset of interband excitations. The individual excitations are labeled by the corresponding
     final-state LL and also by particular color-coding in (d), where the field-dependence of the LL spectrum ($n=0,1,2\ldots$6), calculated for $E_g=-60$~meV and $\delta=15$~meV,
     has been plotted. The velocity parameter was fixed at $v=0.94\times10^6$~m/s and
     the strength of the spin-orbit coupling at $\Delta=400$~meV. The model does not include any other free parameters. The CR data are taken from our previous
     magneto-reflectivity experiments.\cite{AkrapPRL16} The dashed line in (d) denotes an approximate position of the Fermi energy as a function of $B$. In the parts (a-c), the lowest in energy interband transition (light blue) may be identified as the excitation from the flat band to 2$_\downarrow$ level, but its energy also coincides
     with the intraband 1$_\downarrow \rightarrow$3$_\downarrow$ excitation, which may become electric-dipole-active due to the lack of full isotropy in the studied system.
     \label{Bodnar versus data}}
\end{figure*}

Let us first analyze these data in a simplified way, using the semiclassical Bohr-Sommerfeld quantization,\cite{Onsager1952,GoerbigEPL14}
which only considers the orbital motion of electrons and neglects their spin.
In this picture, the density of states, and consequently, also the joint density of states, becomes modulated due to the quantized cyclotron motion of electrons.
With an assumption of isotropic bands (reasonably well justified for Cd$_3$As$_2$\,\cite{AkrapPRL16})
the cyclotron orbits are characterized by the momenta $k=\sqrt{2eB(n+\gamma)/\hbar}$, where $n$ is integer ($n\geq 1$) and $\gamma$ is a factor related to the Berry phase
of the explored charge carriers ($0\leq\gamma<1$).
This allows us to reconstruct the original band structure, \emph{i.e.}, to get the energy distance from the valence to conduction band,
from which the corresponding LLs are formed (the HH band and the upper cone).

Setting the indices $n$ as specified in Fig.~\ref{Fan chart}b,c ($n=2,3,4,5$ and 6), all experimental points reasonably well fall
on a single line (inset of Fig.~\ref{Fan chart}c), thus implying a nearly perfect linear energy-momentum dependence, with the slope reflecting the velocity
parameter $v\approx0.95\times10^6$~m/s. In the inset of Fig.~\ref{Fan chart}c, we have considered, consistently with the Kane/Bodnar model, the phase factor $\gamma=1/2$.
Importantly, the deduced velocity exactly matches the value deduced for the upper conical conduction band in the original analysis of Bodnar,\cite{Bodnar77}
in the more recent STM/STS study\cite{JeonNatureMater14} as well as in cyclotron resonance experiment~\cite{AkrapPRL16}. This remarkable agreement
may only be explained by the existence of a nearly dispersionless valence band, from which electrons are promoted,
via an optical excitation, to the conical conduction band (excitations from the blue to violet band in Fig.~\ref{Scheme}). This is fully consistent
with the Bodnar/Kane model\cite{KaneJPCS57}, which implies the existence of such a flat band. On the other hand, no flat band is expected
in systems described by the Dirac Hamiltonian.

Now we analyze the negative offset of interband inter-LL excitations (seen in Fig.~\ref{Fan chart}) in greater detail. Again, no such
offset is expected for Dirac-type systems, with an exception for the unlikely case that the velocity parameter increases isotropically
with momentum. This would imply strong parabolic corrections to the linear bands and thus equally strong deviations from the $\sqrt{B}$ behavior
of the magneto-optical transmission lines. The absence of such deviations precludes this type of band corrections as the origin of the negative offset.
In contrast, the offset -- either positive or negative -- is expected in Kane semiconductors/semimetals with a non-zero band gap $E_g$,
as visualized using the asymptotes in Fig.~\ref{Energy-momentum}b. At photon energies larger than the band gap, $\hbar\omega\gg E_g$,
the flat band becomes shifted with respect to the upper cone by $E_g/2$, which gives rise to the offset of the interband inter-LL excitations
(when followed as a function of $\sqrt{B}$). The same conclusion can be drawn directly from the LL spectrum of Kane electrons.
At photon energies significantly higher as compared to $E_g$, the energies of flat-to-upper-cone inter-LL transitions read (at $k=0$):
$E_g/2+v\sqrt{2eB\hbar(n-1/2+\sigma)}$ for $n\geq 1$, where $\sigma=\pm1/4$ stands for the spin projection, see Ref.~\onlinecite{OrlitaNaturePhys14} 
and related Supplementary materials.

When the full Bodnar model -- with a non-zero crystal field splitting parameter $\delta$ -- is considered, the situation becomes more complex and
the band structure is no longer isotropic. This is best manifested by the modulation of the HH band, which is completely flat only for $\delta\equiv 0$.
This modulation is illustrated in Figs.~\ref{Energy-momentum}a,c, where the LH, HH and E bands have been plotted for two different crystallographic directions. These
are perpendicular to the magnetic field applied along the [112] direction and they were chosen to visualize the full width of the flat
band induced by $\delta$. For relatively small $\delta$, the HH band remains
fairly flat, but it is shifted by the average value of $\delta/3$. This implies an approximate offset of flat-to-cone inter-LL excitations: $E_g/2+\delta/3$.

To estimate the $E_g$ and $\delta$ parameters, the simplified analysis of our magneto-optical data based
on the semi-classical Bohr-Sommerfeld model (see the inset of Fig.~\ref{Fan chart}c) has to be replaced with full
quantum-mechanical treatment. To this end, we have numerically calculated the LL spectrum within the complete Bodnar model
for the magnetic field oriented along the [112] direction~\cite{SM},\phantom{\cite{Wallacepss79,OrlitaPRL12}} see Fig.~\ref{Bodnar versus data}.
In these calculations, the velocity parameter was considered as isotropic and fixed at $v=0.94\times10^6$~m/s,
which is a value consistent with the original Bodnar's analysis\cite{Bodnar77} as well as with more recent STM/STS and cyclotron resonance
studies.\cite{JeonNatureMater14,AkrapPRL16} The spin-orbit coupling, having rather weak influence on the resulting LL spectrum,
was set as $\Delta=400$~meV.\cite{AkrapPRL16}

The only tunable parameters in the calculations of the LL spectrum in Fig.~\ref{Bodnar versus data} were thus the band
gap $E_g$ and the crystal-field splitting $\delta$. We have considered their various combinations, with the approximate boundary condition
implied by the offset $E_g/2+\delta/3\approx-35$~meV determined and discussed above. This is illustrated in Fig.~\ref{Bodnar versus data},
where our calculations are shown for (a): $E_g=-45$~meV, $\delta=-40$~meV, (b) $E_g=-70$~meV, $\delta=0$~meV and
(c) $E_g=-95$~meV, $\delta=40$~meV, respectively. For electric-dipole-active excitations, which preserve spin and follow
the selection rules $n \pm1 \rightarrow n$,\cite{OrlitaNaturePhys14} we obtain two transitions from the flat band into
a given final-state LL with the index $n_\uparrow$ or $n_\downarrow$, see~Ref.~\onlinecite{SM} for more details.
Experimentally, the splitting of transitions due to spin is resolved only for the lowest final-state LL $n=2$, \emph{cf.}, Fig.~\ref{Bodnar versus data}d.

The best agreement between the experimental data and theoretically expected
energies of inter-LL excitations was found for $|\delta|\approx 0$, for which the Bodnar model reduces down to
even more simple Kane model with an effectively vanishing energy scale of Dirac cones. For the crystal-field splitting as low as $|\delta|=40$~meV, the departure of theoretical
lines from experimental points becomes significant enough that it allows us to set, while being conservative in this estimate,
the interval of acceptable values to $|\delta|<40$~meV.

Let us recall that it is the sign of the $\delta$ parameter, which determines whether the system with an inverted band gap becomes
a 3D Dirac semimetal (for $\delta>0$) or a topological insulator (for $\delta<0$). Clearly, our magneto-optical data do not allow
us to determine the sign of $\delta$. Nevertheless, one may profit, in this case, from ab initio calculations of the Cd$_3$As$_2$ band structure presented, for instance,
in Refs.~\onlinecite{WangPRB13,AliIC14,BorisenkoPRL14,ConteSR17}. The absolute values of energy splitting and band gaps deduced
theoretically may not be very accurate. On the other hand, the ordering of bands,
which among other things implies the sign of $\delta$, represents rather reliable output. The available
theoretical works predict positive value of $\delta$, see, \emph{e.g.}, Refs.~\onlinecite{WangPRB13,ConteSR17}. This allows us to reduce the interval of the
expected crystal field splitting to $0<\delta<40$~meV.

As a matter of fact, it is the tetragonal lattice of Cd$_3$As$_2$ itself, with the out-of-plane lattice constant
elongated with respect to the in-plane components ($c/2>a=b$), which determines the sign of the $\delta$ parameter. In HgTe, which is another
well-known Kane semimetal ($E_g=-300$~meV), lateral expansion of the lattice gives rise to a negative crystal-field splitting ($\delta<0$),
and therefore, to a topological insulating phase, which has been extensively studied experimentally.\cite{BrunePRL11,DziomNatureComm17} In contrast, compressive
lateral strain is expected to transform HgTe into a 3D symmetry-protected Dirac semimetal (with $\delta>0$). Therefore, staying strictly
with estimates based on our experimental magneto-optical data ($|\delta|<40$~meV) only, one cannot unambiguously decide whether Cd$_3$As$_2$ is a symmetry-protected
Dirac semimetal or a topological insulator with a very narrow band gap in bulk.

Let us also note that the negative sign of the band gap in Cd$_3$As$_2$ reflects an inverted ordering of bands in this system,
and consequently, implies the appearance of surface states, similar to those in topological insulators.\cite{HasanRMP10} Such surface states have already
been observed in another inverted-gap Kane semimetal HgTe.\cite{LiuPRB15} In Cd$_3$As$_2$, however, the situation might be more complex. The surface states, which appear
due to the band inversion, might coexist with the Fermi arcs, which are predicted for this material due to the presence of a pair of Dirac nodes
(two pairs of Weyl nodes) at low energies. At present, our magneto-optical data do not provide us with any clear signature of surface states,
but their presence was discussed in the context of ARPES and magneto-transport experiments.\cite{YiSR14,ZhangNC17,SchumannPRL18}

\section{Conclusions}

In summary, we have performed magneto-transmission experiments on Cd$_3$As$_2$. The observed magneto-optical response, comprising a series
of interband inter-Landau level excitations, allow us to determine the band structure parameters relevant for the scale and shape
of possibly present Dirac cones in this material. The estimated crystal field splitting, $0< \delta <40$~meV, represents the very upper limit
for the energy of Dirac electrons in Cd$_3$As$_2$. These relativistic-like particles thus may emerge in the band structure only
at energies far below the Fermi level, which in available samples typically exceeds 100~meV (in our case $E_F\approx160$~meV).
At energies significantly higher than $\delta$, the system behaves as an ordinary Kane semimetal\cite{KaneJPCS57,KacmanPSSB71,CardonaYu,AkrapPRL16}
characterized by a small inverted band gap of $E_g=(-70\pm20)$~meV.

\subsection*{Acknowledgements}
This work was supported by ERC MOMB (No. 320590), TWINFUSYON (No. 692034), Lia TeraMIR, TERASENS, by MEYS CEITEC 2020 (No. LQ1601) and by ANR DIRAC3D projects and MoST-CNRS exchange programme (DIRAC3D). We acknowledge the support of LNCMI-CNRS, a member of the European Magnetic Field Laboratory (EMFL). The authors also acknowledge discussions with D.~M.~Basko, A.~O.~Slobodeniuk and N. Miller. A.~A. acknowledges funding from The Ambizione Fellowship of the Swiss National Science Foundation. I.~C. acknowledges support from the postdoc
mobility programme of the Swiss NSF.


%

\newpage

\thispagestyle{empty}
\clearpage
\addtocounter{page}{-1}
\begin{figure}[htp]
\includegraphics[page=1,trim = 17mm 17mm 17mm 17mm, width=1.0\textwidth,height=1.0\textheight]{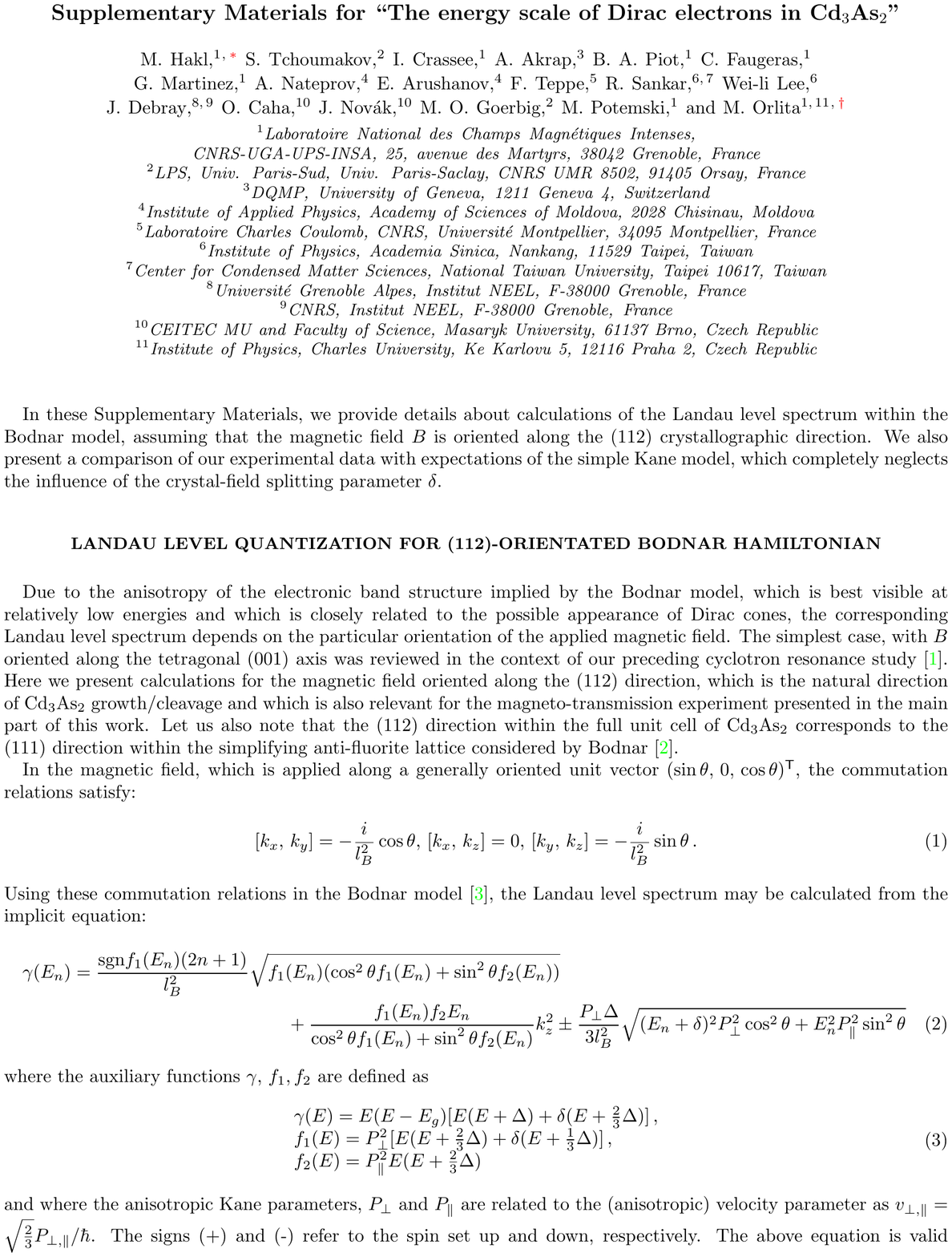}
\end{figure}

\newpage

\thispagestyle{empty}
\clearpage
\addtocounter{page}{-1}
\begin{figure}[htp]
  \includegraphics[page=2,trim = 17mm 17mm 17mm 17mm, width=1.0\textwidth,height=1.0\textheight]{SupplementaryMaterials.pdf}
\end{figure}

\newpage

\thispagestyle{empty}
\clearpage
\addtocounter{page}{-1}
\begin{figure}[htp]
  \includegraphics[page=3,trim = 17mm 17mm 17mm 17mm, width=1.0\textwidth,height=1.0\textheight]{SupplementaryMaterials.pdf}
\end{figure}

\newpage

\thispagestyle{empty}
\clearpage
\addtocounter{page}{-1}
\begin{figure}[htp]
  \includegraphics[page=4,trim = 17mm 17mm 17mm 17mm, width=1.0\textwidth,height=1.0\textheight]{SupplementaryMaterials.pdf}
\end{figure}

\newpage

\thispagestyle{empty}
\clearpage
\addtocounter{page}{-1}
\begin{figure}[htp]
  \includegraphics[page=5,trim = 17mm 17mm 17mm 17mm, width=1.0\textwidth,height=1.0\textheight]{SupplementaryMaterials.pdf}
\end{figure}

\thispagestyle{empty}
\clearpage
\addtocounter{page}{-1}
\end{document}